\documentclass[12pt]{article}
\usepackage{graphicx}

\usepackage{epstopdf}
\begin{document}

\title{Interpretation of the small grains in the inclusions of ice cores}
 
\author{W. Baltensperger\footnote{Centro Brasileiro de Pesquisas F\'\i sicas, Rua Dr.\thinspace Xavier Sigaud,150, 222\thinspace 90 Rio de Janeiro, Brazil;\qquad \qquad e-mail: baltens@cbpf.br}\, and W. Woelfli\footnote{ Institute for Particle Physics, ETHZ H\"onggerberg, CH-8093 Z\"urich, Switzerland (Prof.~emerit.); e-mail: woelfli@phys.ethz.ch .}}
\date{ \today}
\maketitle

\abstract\noindent{\emph{The origin of the grains with diameters less than 4 $\mu$m from the inclusions in Antarctic ice cores  is discussed. It is proposed that these grains were initiated by ions stopped in the upper atmosphere. The ions form molecules, which coagulate and diffuse downwards. These processes give rise to a characteristic  mass spectrum of the small grains. Inclusions in ice were abundant during cold periods of the Pleistocene. This supports a model in which the influx of particles is large when  Earth's orbit stays within a disk shaped cloud around the Sun.  The production of such a cloud may favor light atoms. This should e.g.\ be apparent in the isotope distribution of Magnesium from the small grains. The large grains, which have a variable mass distribution, are terrestrial.}}
\maketitle

\section{Revealing  facts}
Cold periods of the Pleistocene ice ages are correlated with large quantities of inclusions in the ice. A survey of Antarctic ice cores showed that in cold periods the masses are larger by one to two orders of magnitude compared to the values of warm periods \cite{EPICA800,Lambert}. An analysis with detailed  time resolution correlates quantitatively  low temperatures  with high impurity content \cite{Lambert}. The distribution of mass of the dust  versus grain size shows a clear distinction between small and large grains, where the boundary is at a diameter of about 4 $\mu$m \cite{Steffensen,Delmonte}. The terrestrial origin of the large grains has been unambiguously determined by mineralogical methods \cite{Biscaye}. For the large grains the mass distribution versus grain size is irregular and varies from one period to the next  \cite{Steffensen}. In contrast, the small grains show a bell shaped mass distribution, which can be easily parametrized, and which varies little between cold periods. In order to determine the origin of the small grains the  isotope distributions of Sr and Nd has been compared with that of samples from many regions of the globe  \cite{Delmonte}. An agreement was found between material from  Antarctica and from Patagonia.  In Table 1b of \cite{Delmonte}  the samples of South America  are dated; those of the Pampas  have ages between 10 and 25 kyr BP, while the samples from other regions are without dates. It was concluded that the small grains have been  transported from Patagonia to Antactica  \cite{Delmonte,Gaiero}.  However, it is also possible that the grains from the two regions  have the same extraterrestrial origin. 

\section{Extraterrestrial atoms}
The 100 ka period is a dominant feature of the Pleistocene ice ages. While the ellipticity of Earth's orbit has a 100 ka periodicity, the corresponding variation of the insolation is minute, so that it would require a large amplification to explain the observed temperature variations of the period. 
R.A. Muller and G.J. McDonald \cite{Muller1,Muller2} pointed out that the inclination of Earth's orbit  (the angle between Earth's orbital plane and the invariant plane of the planetary system, which is perpendicular to its angular momentum), also has an approximate 100 ka period. They postulated a disk shaped cloud circling the Sun. Then a change of angle between Earth's orbital plane and the plane of the cloud modulates the solar irradiation on the Earth. The cloud  would also produce a correlation between cold periods and impurity content in ice cores \cite{Lambert}, provided that the cloud extends beyond Earth's orbit.  In a Milankowich theory without the cloud the inclination is irrelevant, since the Sun radiates isotropically. 

Woelfli et al.\,\cite{latitude} focussed their attention on another unexplained fact, namely the existence of remains of Mammoths deep in Arctic East Siberia. It indicates that this region had a lower latitude in the Pleistocene. The authors developed a model for a rapid geographic shift of the poles. This shift terminates the Pleistocene. As a by-product of the model a cloud of atoms and ions circling the Sun had to exist during a time of order of a few million years. This agrees with the observed total length of time of the Pleistocene ice ages  \cite{Tiedemann}, i.e.\ about 2.5 to 3 Ma. In contrast the Milankowich theory has no time limit backward or forward.

\section{Mass distribution of the grains}
The  extraterrestrial ions and atoms enter into the high atmosphere with planetary velocity. They are stopped and form molecules. These diffuse and coagulate with others. Clusters move downwards, occasionally joining others until finally they reach the ground. 

As a didactic illustration we imitated such a process in a simple computer model. It contains   six vertical layers of a two dimensional square net with 8 lattice points on each horizontal line. The horizontal lines are closed to a circle to avoid boundaries. The top layer receives particles in random positions, which combine with any cluster that may exist there. 
Then in a random position of the net, any cluster that may exist there makes a side step with a probability that decreases with cluster size. It combines with a cluster that it may encounter. Finally in random positions of the net the clusters make a downward step with a probability that increases with cluster size, also combining with any cluster they may find. This sequence is repeated a million times. Fig.\ 1 shows a plot of the mass that reaches the ground layer for each cluster size. The fluctuations vary from run to run. Consistently the curve shows a bell shape.   Smaller  than typical clusters are rare, because small clusters sink slowly and have time to grow before they reach the ground. Similarly  large clusters diffuse downwards with good speed and have no time to grow to excessive size.  

We do not intend to make this simple program look more realistic, since this could only show once more that in a complex model  open parameters can be ajusted to fit experimental data. A determination of realistic parameters from the chemistry of the atmosphere is beyond our capabilities. The purpose of our simple program is merely to make a bell shaped mass distribution plausible. 

\section{Proposed experimental test}
In the model of the Pleistocene ice ages of Woelfli et al.\,\cite{latitude} the atoms or ions of the disk-shaped cloud around the Sun had been emitted from a planetary object in an extremely excentric orbit. This object was hot from tidal work and solar radiation. Particle emission from it is limited by the escape velocity. This favours light particles and leads to an isotope effect  within an atomic species. Magnesium which has three stable isotopes should be a good candidate for a test.  Being light it is easily emitted, so that extraterrestrial atoms would outnumber those of terrestrial origin.  A measurement of the isotopic distribution of Mg from small clusters of a cold period could reveal its extraterrestrial origin.

\begin{figure}[htbp]
\begin{center}
\includegraphics[width=4in]{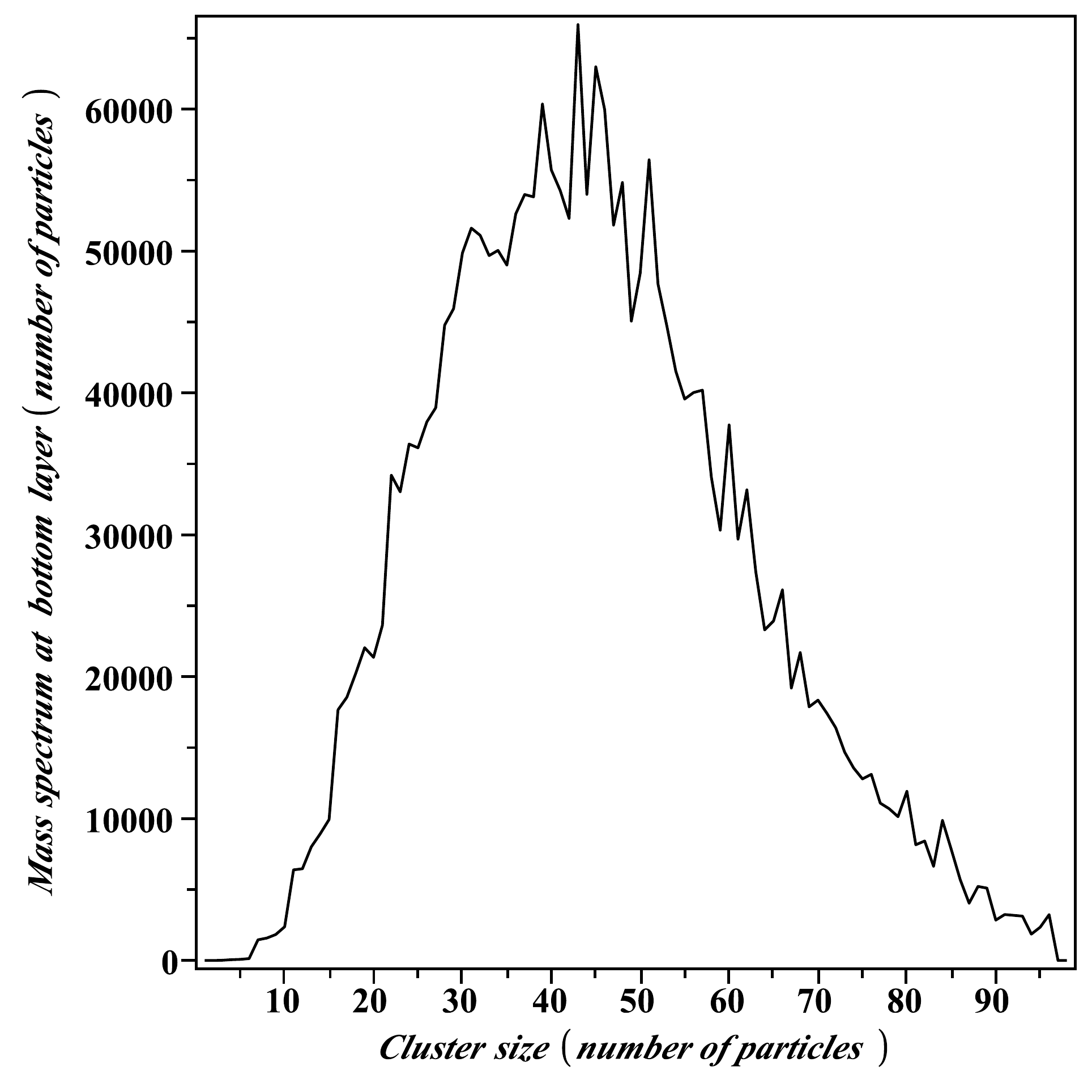}
\caption{{\bf Mass distribution versus cluster size in a didactic model.}}
\end{center}
\end{figure}

\end{document}